\documentclass[superscriptaddress,letterpaper, aps, amssymb, preprint, amsmath, 12pt]{revtex4-1}
\usepackage[justification=raggedright,singlelinecheck=false,labelfont=bf,labelsep=period]{caption,subcaption}

\usepackage{xr}
\usepackage{braket}

\usepackage{siunitx}

\usepackage{comment}

\usepackage{nameref}

\newcommand{\wmk}{Wm$^{-1}$K$^{-1}$}

\newcommand{\SM}{Supporting Information}

\usepackage{scalerel}

\usepackage{tikz}

\usetikzlibrary{svg.path}
\definecolor{orcidlogocol}{HTML}{A6CE39}
\tikzset{
  orcidlogo/.pic={
    \fill[orcidlogocol] svg{M256,128c0,70.7-57.3,128-128,128C57.3,256,0,198.7,0,128C0,57.3,57.3,0,128,0C198.7,0,256,57.3,256,128z};
    \fill[white] svg{M86.3,186.2H70.9V79.1h15.4v48.4V186.2z}
                 svg{M108.9,79.1h41.6c39.6,0,57,28.3,57,53.6c0,27.5-21.5,53.6-56.8,53.6h-41.8V79.1z M124.3,172.4h24.5c34.9,0,42.9-26.5,42.9-39.7c0-21.5-13.7-39.7-43.7-39.7h-23.7V172.4z}
                 svg{M88.7,56.8c0,5.5-4.5,10.1-10.1,10.1c-5.6,0-10.1-4.6-10.1-10.1c0-5.6,4.5-10.1,10.1-10.1C84.2,46.7,88.7,51.3,88.7,56.8z};}}
\newcommand\orcidicon[1]{\href{https://orcid.org/#1}{\mbox{\scalerel*{
\begin{tikzpicture}[yscale=-1,transform shape]
\pic{orcidlogo};
\end{tikzpicture}
}{|}}}}

\usepackage{etoolbox}

\makeatletter
\pretocmd\frontmatter@thefootnote{\color{blue}}{}{}

\usepackage{dcolumn} 

\usepackage{titlesec}
\titlespacing*{\section}
{0pt}{4ex}{1.3ex}
\titlespacing*{\subsection}
{0pt}{4ex}{0ex}

\titleformat{\section}
  {\bfseries\MakeUppercase}{\thesection\raggedright}{1em}{}
  \titleformat{\subsection}
 {\bfseries}{\thesubsection\raggedright}{1em}{}

\usepackage{caption}
\usepackage{xstring}
\usepackage{xspace}
\usepackage[utf8]{inputenc}
\usepackage{natbib}
\usepackage{gensymb}
\usepackage{mathtools}

\usepackage{float}
\usepackage{graphicx}

\usepackage{url}
\usepackage[utf8]{inputenc}
\usepackage[T1]{fontenc}
\usepackage{textcomp}
\usepackage{gensymb}
\usepackage[colorlinks,citecolor=red,urlcolor=blue,bookmarks=false,hypertexnames=true]{hyperref}
\usepackage{booktabs}
\usepackage{multirow}
\usepackage{textcomp}
\usepackage{graphicx}

\begin{document}

\title{High thermal conductivity dominated by thermal phonons with mean free paths exceeding hundred nanometer in three-dimensional covalent organic framework derivatives: a molecular dynamics study}

\author{Sungjae Kim~\orcidicon{0009-0000-4062-1176}}

\affiliation{%
Department of Mechanical Engineering, Seoul National University, Seoul 08826, Republic of Korea}%

\author{Taeyong Kim~\orcidicon{0000-0003-2452-1065} }
\email{tkkim@snu.ac.kr}

\affiliation{%
Department of Mechanical Engineering, Seoul National University, Seoul 08826, Republic of Korea}%
\affiliation{
Institute of Advanced Machines and Design, Seoul National University, Seoul 08826, Republic of Korea
}
\affiliation{%
Inter-University Semiconductor Research Center, Seoul National University, Seoul 08826, Republic of Korea}%

\date{\today}

\begin{abstract}

Thermal properties of covalent organic frameworks (COFs) are of fundamental interest owing to exceptional heat conduction properties. Recent studies have suggested that their thermal conductivities can be enhanced by multiple factors such as pore size, mass density, and degree of chain order. However, microscopic processes that govern heat conduction properties have been explored in only a limited number of COFs. Here, we report thermal transport properties of 3D COF derivatives using molecular dynamics simulations. In this work, we have studied six different COF-102 derivatives with different organic linkers and topologies. Among the derivatives studied, we found that COF-102 derivatives with high mass density can exhibit thermal conductivity as high as $\sim 27$ \wmk, owing to suppressed chain rotation that leads to thermal phonons scattered by anharmonicity. Our results show that the observed orders of magnitude of increase in the thermal conductivity are primarily attributed to low frequency phonon modes that support hundreds of nanometer scale mean free path, which predominantly carry heat. Our study provides a theoretical framework that elucidates the structure-property relationship governing heat conduction in COFs, offering design strategies for thermal management applications.

\end{abstract}

{\global\let\newpagegood\newpage\global\let\newpage\relax\maketitle}

\newpage
\section{Introduction}    

Covalent organic frameworks (COFs), a class of crystalline porous organic materials, hold promise for various applications such as gas storage and separation \cite{furukawa_appliedgas_2009,appli_gasse_2018}, catalysis \cite{appli_cata_2018}, optoelectronics~\cite{keller_optoelec_2021} and energy storage \cite{zhu_battery_2021}, owing to their large surface areas, tunable pore sizes, and high thermal stabilities \cite{Yaghi_first2D_2005,Yaghi_first3D_2007,Yaghi_science_2017}. Prior extensive studies have suggested that the diverse nature of structure, composition, and functionalities facilitates the tunability of their physical properties for tailored applications \cite{huang_platform_2016, liu_platform2_2021}.
For instance, conjugated COFs have shown unique optoelectronic phenomena such as exceptionally high charge carrier mobility \cite{guo_piconjugated_photodetector_2013} and tunable fluorescence emission as a result of cation addition \cite{ding_piconjugated_sensing_2016}. Additionally, COFs could exhibit other interesting properties such as facile ion transport \cite{wang_appli_redox_2017,li_review_redox_active_2021} and thermoelectricity \cite{wu_TE_DFT_2023,zhou_TE_2023}, which are crucial for energy storage and conversion applications~\cite{RunHu_PRX_ML}. However, efficient heat dissipation is needed in such applications to ensure device stability \cite{HeatDissi_NSR}, making it essential to understand heat transport properties in COFs \cite{shakouri_chiphotspot_2006,Ericpop_chipheat_2010, babaei_MOFadsorption_nature_2020}.

Thermal conduction in COFs is interesting due to several reasons. First, bulk COFs both in crystalline and amorphous phases exhibit exceptionally low thermal conductivities close to those of many bulk polymers. While lacking existing literature, amorphous COFs, including COF-300 and reticular innovative organic materials (RIO) derivatives, were reported to have thermal conductivities of $\sim 0.05$ \wmk~\cite{freitas_example_poresize_2017} that do not exhibit marked differences with those of crystalline COFs \cite{thakur_interpen_2023,liu_example_boronk_2016}.
Second, due to their inherent porosity, COFs generally show a negative correlation between thermal conductivity and pore size, despite few exceptions \cite{ma_example_ultrahigh_2021,giri_example_2Dgask_2021}. Given the structural tunability of COFs, it has been suggested that their thermal transport properties can be modulated through rational design of linkers and net topologies.
Third, constituent light elements (e.g., B, C, N) connected by strong covalent bonds offer possibilities of enhancing heat conduction properties. In fact, prior studies have observed that heat conduction in crystalline porous materials becomes increasingly impeded as linker-node mass mismatch gets larger \cite{islamov_bottleneck2_2023}. A representative example is metal organic frameworks (MOFs), which unavoidably contain metal nodes that exhibit a large mass mismatch with organic linkers. This leads to additional vibrational scattering, which, in turn, suppresses thermal conductivity \cite{wieser_bottleneck1_2021,zhou_bottleneck2_2022}. 
In support of these observations, while MOFs generally exhibit glass-like thermal conductivities typically less than $\sim 0.8$ \wmk~\cite{wang_MOF_74_2015,zhang_MOF_5_2021,huang_UIOpowder_2019,babaei_MOFadsorption_nature_2020}, recent experimental work for 2D crystalline COF thin films has reported thermal conductivity exceeding 1 \wmk~along with high longitudinal speed of sound \cite{evans_naturematerial_2021}.

COFs share chemical and structural similarities with polymers, suggesting their potential for enhancing thermal transport properties.
According to prior studies, polymers can be thermally conductive if they have a large atomic mass density \cite{xie_blends_density_2016,xie_amor_density_2017}, which is associated with densely packed crystalline domains composed of aligned covalent chains \cite{lv_liquid_density_2022}. In the extreme, perfect polymer crystals, such as polyethylene (PE) \cite{shulumba_PRL_2017} and polythiophene (PT) \cite{cheng_ab_PT_2019}, have been reported to exhibit thermal conductivities exceeding $100$ \wmk. 
Similarly, in COFs, design strategies such as tuning functional groups and pore sizes to increase mass density have been shown to enhance thermal conductivity. Recently, Rahman et al. observed that thermal conductivity of 2D COFs monotonically increases with mass density~\cite{rahma_example_engineering_2023}. The observed trend was more pronounced in COFs with nodes that have increased connectivity through stronger covalent bonds to the surrounding linkers, resulting in weak anharmonicity.
Additionally, Ma et al. demonstrated that COF-300 derivatives can achieve high thermal conductivity of $\gtrsim 15$ \wmk~along all spatial directions when pore size is reduced to 0.63 nm~\cite{ma_example_ultrahigh_2021}. In support of these findings, a recent high-throughput screening study that surveyed over 10,000 3D COFs reported that thermal conductivity as high as $\sim 50$ \wmk~can be achieved through enhanced chain orientations and appropriate choice of linkage material \cite{thakur_example_highthrou_2024}. These findings suggest that understanding of these design factors could improve the intrinsic upper limit of thermal conductivities in COFs.
 
Here, we investigate the thermal transport properties of COF-102 derivatives using molecular dynamics (MD) simulations. In this work, we have studied six different COF-102 derivatives with different organic linkers and topologies, which vary elastic stiffness, mass density, and pore size in them. Among the derivatives investigated in this work, we found that COF-102-1 (ctn) that has the highest mass density and speed of sound, can exhibit thermal conductivity as high as $\sim 27$ \wmk, nearly two-fold increase compared to the original COF-102, primarily due to suppressed chain rotations, leading to heat-carrying phonons scattered by anharmonicity. The observed increase in the thermal conductivity is primarily due to low frequency phonons with mean free paths (MFPs) on the order of hundreds of nanometers, which are dominant contributors to heat conduction in them. Our study provides a theoretical framework for tailoring thermal properties of 3D COFs for thermal management applications.

\section{Methods} 
\label{sec:method}

We first constructed atomic structures of COF-102 derivatives and optimized their energies using the polymer consistent force field (PCFF) \cite{sun_PCFF_1994} in Large-scale Atomic/Molecular Massively Parallel Simulator (LAMMPS)~\cite{plimpton_LAMMPS_1995}. The detailed structural and physical properties of COF-102 derivatives are provided in \SM~Sec.~SI. Further MD calculations with periodic boundary conditions (PBC) applied along all spatial directions were conducted using PCFF potential in LAMMPS.

\begin{table}[b!]
\centering
\caption{Number of unit cells in each supercell, along with the corresponding domain size used in EMD simulations of COF-102 derivatives.}
\setlength{\tabcolsep}{3.5pt} 
\renewcommand{\arraystretch}{0.5} 
{\small
\begin{tabular}{cccc}
\toprule
\textbf{Material} & \textbf{Topology} & \textbf{Number of unit cells} & \textbf{Supercell size (nm)} \\
\midrule
\multirow{2}{*}{COF-102-1} 
  & ctn & $4 \times 4 \times 4$ & 4.45 \\
  & bor & $6 \times 6 \times 6$ & 4.27 \\
\midrule
\multirow{2}{*}{COF-102-2} 
  & ctn & $2 \times 2 \times 2$ & 5.44 \\
  & bor & $3 \times 3 \times 3$ & 5.28 \\
\midrule
\multirow{2}{*}{COF-102-3} 
  & ctn & $2 \times 2 \times 2$ & 8.59 \\
  & bor & $3 \times 3 \times 3$ & 8.43 \\
\bottomrule
\end{tabular}
}
\label{Tab:EMDsize}
\end{table} 

For EMD simulations, the thermal conductivity $\kappa$ was computed using the Green-Kubo formalism \cite{chen_book_2005}. 
\begin{equation} 
\kappa(T)= \frac{1}{3Vk_BT^2} \int_{0}^{\infty} \langle \mathbf{J}(t) \cdot \mathbf{J}(0) \rangle \, dt
\label{eq:GK}
\end{equation}
Here, $V$ is system volume, $k_B$ is the Boltzmann constant, $T$ is temperature, and $\mathbf{J}$ is heat current. The term $\langle \mathbf{J}(t) \cdot \mathbf{J}(0) \rangle$ represents ensemble average of heat current autocorrelation function (HCACF) that decays with time. Thermal conductivity was calculated by integrating HCACF over a correlation time that is sufficiently longer than time at which HCACF decays to zero~(see \SM~Sec.~SII for details). 

Simulated systems were first relaxed in the NPT, NVT, then NVE ensemble at 300 K for 1 ns. Then, heat current data were collected in the additional NVE ensemble for 5 ns. To reduce statistical errors, thermal conductivities were averaged over eight independent simulation runs that started with different initial velocities. In all EMD simulations, a time step of 1 fs was used. The sizes of simulated structures are summarized in Tab.~\ref{Tab:EMDsize}.

Phonon dispersion relations were calculated by using SED \cite{thomas_SED_2010} as a function of angular frequency $\omega$ and wave vector $\mathbf{k}$, which can be expressed as
\begin{equation} 
\Phi(\mathbf{k}, \omega) = \frac{1}{4 \pi \tau_0 N_T} \sum_{\alpha} \sum_{b}^{B} m_b 
\left| 
\int_{0}^{\tau_0} \sum_{n}^{N_T} \dot{u}_{\alpha} 
\left(
\begin{array}{c}
n \\
b
\end{array};t \right)
\exp \left( i \mathbf{k} \cdot \mathbf{r}_n
 - i \omega t \right)
dt 
\right|^2,
\label{eq:SED}
\end{equation}
where $\tau_0$ is total integration time, $N_T$ is total number of unit cells, $m_b$ and $\dot{u}$ are mass and velocity of atom with index $b$, and $\mathbf{r}_n$ is equilibrium position of each unit cell.

In our SED analysis, we used systems with $n \times n \times 80 $, in which $n$ corresponds to the number of unit cells used in EMD along one spatial direction; the systems underwent identical relaxation process employed in EMD simulations at 300 K, except the additional subsequent NVE ensemble at which velocities and positions of atoms were recorded for 4 ns to calculate the SED. The resulting SED values were averaged over two independent runs. Phonon properties were computed by fitting frequency spread for each phonon mode to Lorentzian function
\begin{equation} 
\Phi(\mathbf{k}, \omega) = \frac{I}{1+[(\omega-\omega_c)/\gamma(\mathbf{k})]^2},
\label{eq:Lorentz}
\end{equation}
where $I$ is peak intensity, $\omega_c$ is frequency at peak position, and $\gamma$ is half-width at half-maximum. As established in the literature~\cite{thomas_SED_2010}, MFPs of phonon modes were determined by $l(\mathbf{k}) = \tau(\mathbf{k}) \cdot v_g(\mathbf{k})$, where phonon lifetime is given by $\tau(\mathbf{k}) = 1/2\gamma(\mathbf{k})$ and phonon group velocity is defined as $v_g(\mathbf{k}) = \partial \omega / \partial \mathbf{k}$.

In NEMD simulations, we varied domain size along the axial direction ($z$) from 33.1 to 198.3 nm, while keeping lateral dimensions equivalent to those used in EMD simulations. The thermal conductivity was computed by using Fourier's law~\cite{Liu_AI, Chakraborty_ACSAMI_2020}:
\begin{equation}  
\kappa = \frac{Q}{\nabla T}
\label{eq:fourier}
\end{equation}
where $Q$ is heat flux and $\nabla T$ is temperature gradient. System was first relaxed in the NPT ensemble at 300 K for 1 ns. Subsequently, two ends with a length of 2.2 nm were set as fixed regions. Langevin thermostats were applied to heat source and sink regions (length: 8.8 nm) to generate steady-state heat flux along the axial direction under a temperature gradient for 4 ns; the data (temperature gradient and heat flux) obtained after initial 1 ns were used to compute the thermal conductivity. Temperatures of heat source and sink regions placed next to each fixed region were set to be at 330 K and 270 K, respectively. A time step of 0.25 fs was used in our entire NEMD calculations.

\section{Results and discussion}

\begin{figure}
{\includegraphics[width=16cm,keepaspectratio]{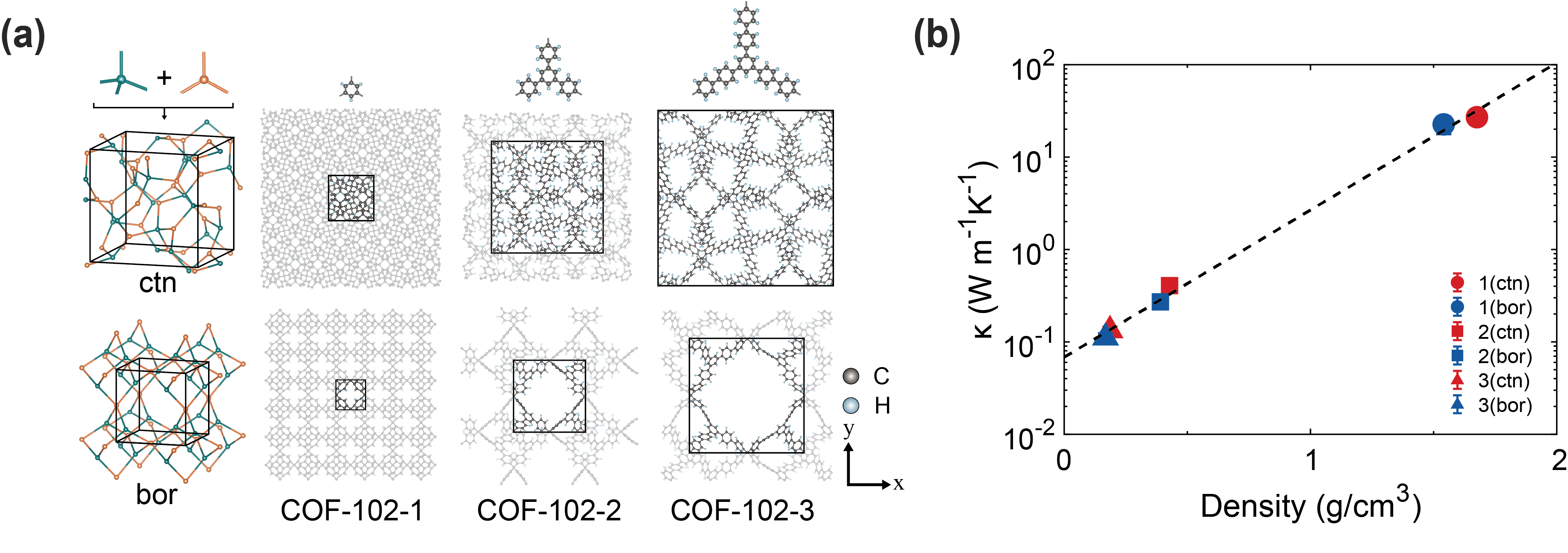}
\captionsetup{justification=justified, singlelinecheck=false}}
    \caption{\label{fig:fig1}
(a) Schematic illustration of topological nets and the corresponding crystal structures of COF-102 derivatives. Each structure exhibits a distinct unit cell (rectangular box with solid lines) and pore size. The lattice constants ($a=b=c$) for each structure are given by: COF-102-1 (ctn: 11.01 \text{\r{A}}; bor: 7.13 \text{\r{A}}); COF-102-2 (ctn: 26.87 \text{\r{A}}; bor: 17.44 \text{\r{A}}), and COF-102-3 (ctn: 42.59 \text{\r{A}}; bor: 27.73 \text{\r{A}}).
The three phenyl-based linkers shown at the top represent the 3-connected linkers of each framework (orange in the left panel), which are connected to tetrahedral carbon nodes (green in the left panel). 
(b) Calculated thermal conductivities of COF-102 derivatives (red: ctn ; blue: bor) versus mass density. Each circle, square and triangle correspond to the COF-102-1, COF-102-2, and COF-102-3, respectively. The dashed line indicates the trend of increasing thermal conductivity with increasing mass density. Maximum thermal conductivity of $\sim 27$~\wmk~was observed in the COF-102-1 (ctn) structure. We note that while we included the error bars representing 95\% confidence interval determined from eight independent simulations, they are not clearly visible as they are smaller than the symbols.}
\end{figure}

We studied the thermal transport of COF-102 derivatives using MD simulations. Detailed descriptions of the MD simulations implemented in this work are given in \SM~Sec.~SII and \nameref{sec:method}. 
As illustrated in Fig.~\ref{fig:fig1}(a), the COF-102 derivatives studied in this work were constructed from two cubic topologies, carbon nitride (ctn) and boracite (bor). They are formed by linking tetrahedral carbon nodes with 3-connected linkers; these nodes are connected to three distinct linkers via C-C bonds. The size of linker depends on the number of phenyl rings it contains. Although both ctn and bor topologies share the same stoichiometry, bor topologies are generally $\sim 15$\% less dense than ctn topologies (see \SM~Sec.~SI for details).

Figure~\ref{fig:fig1}(b) shows room temperature thermal conductivities of COF-102 derivatives, calculated using the Green-Kubo formalism (Eq.~\ref{eq:GK}). As expected, a positive correlation between thermal conductivity and mass density is clearly visible. We note that the opposite trend is observed between thermal conductivity and pore size (see \SM~Fig. S3 in Sec.~SII). COF-102-1, in particular, exhibits the highest thermal conductivity ($\kappa_{ctn}\sim27$ \wmk; $\kappa_{bor}\sim22$ \wmk), while the others show thermal conductivities reduced by nearly two orders of magnitude. In all cases, bor topologies yield $\sim 20$\% lower thermal conductivities than ctn topologies. We here note that COF-102-2 (ctn) is structurally similar to the original COF-102, except that its linkers are formed by substituting all atoms in the boroxine (B$_3$O$_3$) rings with carbon atoms. This leads to enhanced thermal conductivity of $\sim 0.4$ \wmk, which is $\sim30$\% larger than the reported value of pristine COF-102 ($\sim 0.3$ \wmk~\cite{liu_example_boronk_2016}). According to the literature, low thermal conductivity of COF-102 primarily originates from vibrational mismatches between B and O atoms~\cite{liu_example_boronk_2016}. In contrast, such mismatches can be avoided in our derivatives due to the presence of all-carbon linkers.
 
\begin{figure}
 {\includegraphics[width=16cm,keepaspectratio]{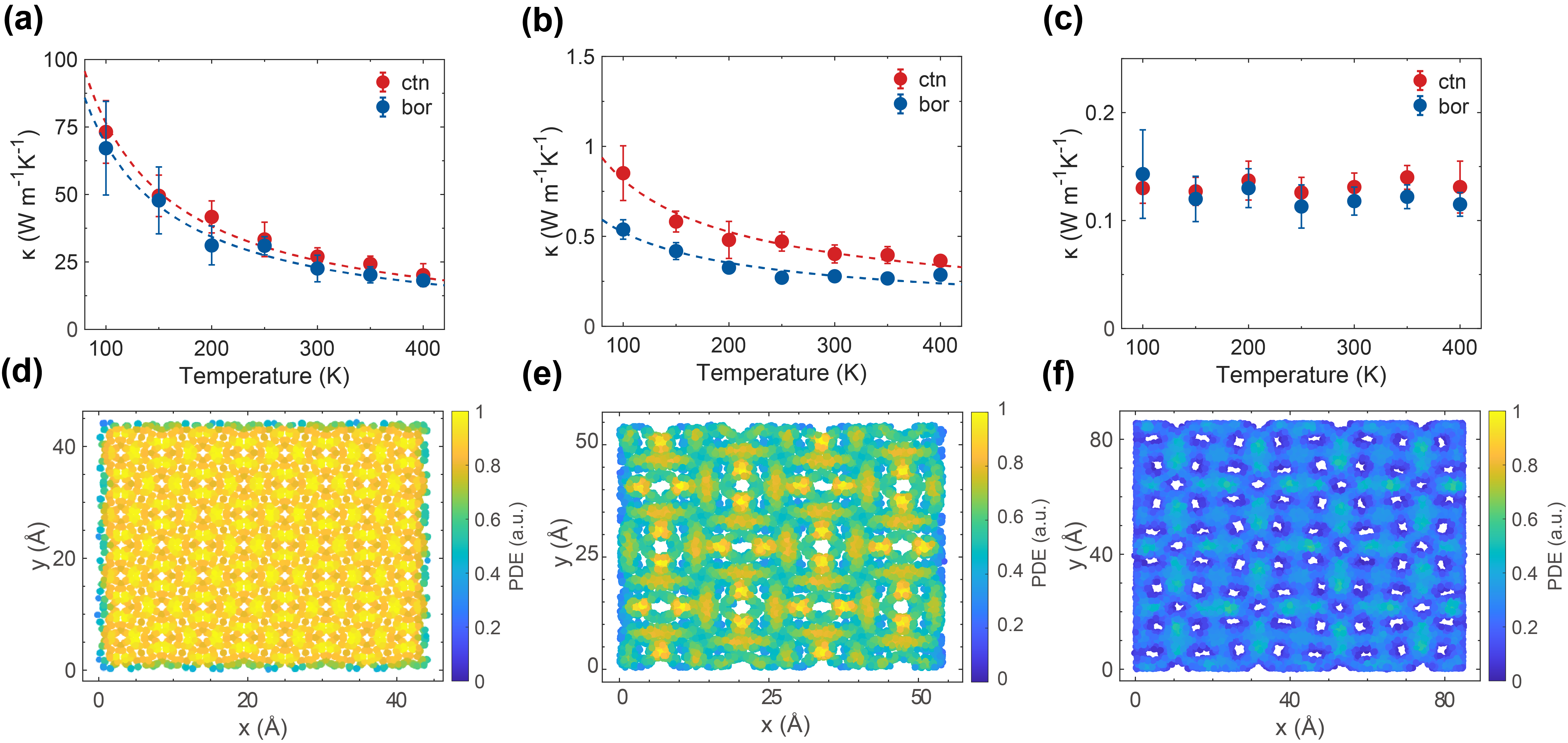}}
    \caption{\label{fig:fig2}
Thermal conductivity versus temperature ranging from 100 K to 400 K for (a) COF-102-1, (b) COF-102-2, and (c) COF-102-3. The thermal conductivity of COF-102-1 exhibits a marked temperature dependence, while that of COF-102-2 and COF-102-3 shows weaker dependence. Assuming power law dependencies $\kappa\propto T^{-n}$, best fit (dashed lines) was obtained as $n \sim 1$ for COF-102-1, and $n \sim 0.6$ for COF-102-2. Also shown are normalized PDE maps for simulated structures of (d) COF-102-1 (ctn), (e) COF-102-2 (ctn), and (f) COF-102-3 (ctn), respectively. The PDE values represent the likelihood of atoms occupying specific positions when projected onto the $xy$ plane. These values are generally high in COF-102-1, intermediate in COF-102-2, and low in COF-102-3.}
\end{figure} 

We also calculated the temperature-dependent thermal conductivity of COF-102 derivatives for temperatures ranging from 100 to 400 K. Results are given in Fig.~\ref{fig:fig2}(a)--(c).
In COF-102-1, thermal conductivity decreases monotonically with increasing temperature, exhibiting a $T^{-1}$ dependence over the entire temperature range considered here. This trend suggests that thermal phonons are predominantly scattered by anharmonicity, as is widely observed in many crystalline solids \cite{chen_book_2005}. On the other hand, COF-102-2 and COF-102-3 exhibit a weak temperature dependence that deviates from classical $T^{-1}$ behavior. In particular, the trend observed in COF-102-3 more closely resembles that of amorphous materials, although calculated X-ray diffraction patterns show distinct peaks at specific angles, indicating a high degree of crystallinity (see \SM~Sec. SIII) \cite{robbins_PCFF_crystalline_2015}.

Following Refs.~\cite{ma_PCFF_chainrotation_2019,ma_example_ultrahigh_2021,hu_PCFF_HOF_2024}, to identify origin of the observed behavior, we computed the probability density estimate (PDE) projected onto the $xy$ plane. PDE is a measure of spatial probability of finding atoms at given spatial locations by projecting all atomic coordinates onto a chosen plane. 
As shown in Fig.~\ref{fig:fig2}(d), COF-102-1 exhibits uniform pore geometries with a high PDE (yellow). In contrast, Figs.~\ref{fig:fig2}(e) and \ref{fig:fig2}(f) show that the PDE progressively decreases (shifting toward blue) from COF-102-2 to COF-102-3, indicating increased pore irregularity compared to COF-102-1. We note that PDE maps for bor topologies exhibit similar trends across the derivatives, as compared to ctn topologies (See \SM~Sec.~SIV for additional PDE maps of bor structures).
Prior studies have established that lower PDE values are associated with greater chain rotation, which reduces overlap of periodic atomic positions, and that such rotational disorder introduces additional vibrational scattering, leading to a weaker temperature dependence in thermal conductivity \cite{ma_PCFF_chainrotation_2019,ma_example_ultrahigh_2021}.
Based on these findings, we suggest that the reduced PDE in low-density COFs could be attributed to increased linker rotations, which contribute to the anomalous temperature dependence observed in these crystalline structures.

\begin{figure}
 {\includegraphics[width=16cm, height=12cm,keepaspectratio]{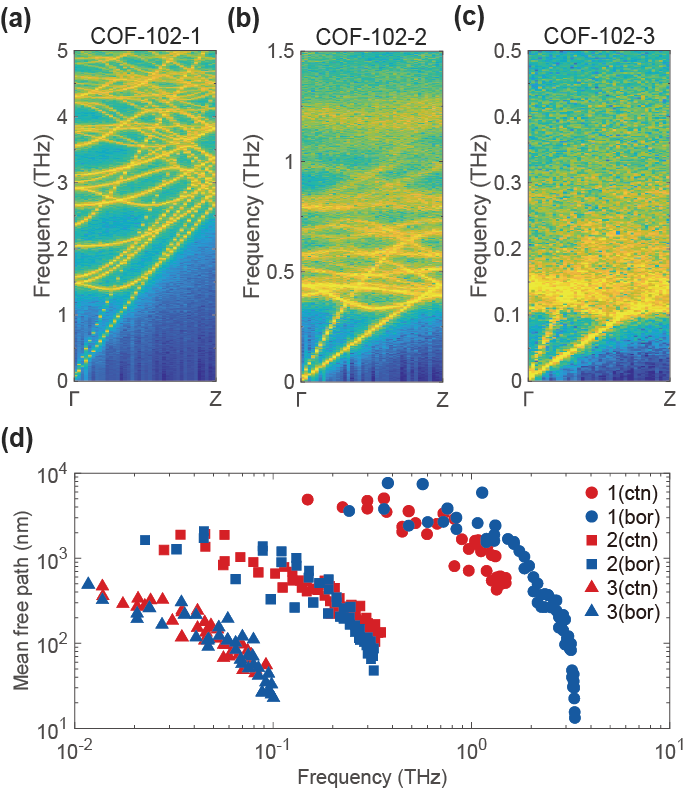}}
    \caption{\label{fig:fig3}
    Low energy SED that represents phonon dispersion relations at 300 K for (a) COF-102-1 (ctn), (b) COF-102-2 (ctn), (c) COF-102-3 (ctn). Full dispersions are given in~\SM~Sec. SV. Acoustic modes are clearly seen for COF-102-1 (ctn) with a stiff LA polarization. (d) MFP versus frequency for acoustic polarizations for each structure. Data was determined from lifetimes yielded by a Lorentzian line-shape fit that shows the best agreement with SED at a certain wave-vector. Sub-THz vibrational modes in COF-102-1 exhibit MFPs exceeding $\sim1$ $\mu$m, while corresponding frequencies of modes occur below a few GHz to several tens of GHz in COF-102-2 and COF-102-3.}
\end{figure}

To gain insight into microscopic details, we calculated spectral energy density (SED) using Eq.~\ref{eq:SED} to extract phonon dispersion and spectral phonon MFPs~\cite{thomas_SED_2010}. Phonon dispersions over selected frequency ranges are shown in Fig.~\ref{fig:fig3}, while full dispersions are given in \SM~Sec.~SV.
As shown in Fig.~\ref{fig:fig3}(a), COF-102-1 exhibits well-defined phonon modes up to 5 THz, indicating the presence of long-lived thermal phonons. Calculated group velocities near the $\Gamma$ point of the longitudinal acoustic (LA) and transverse acoustic (TA) branches are $v_{LA}\sim$ 10 km s$^{-1}$ and $v_{TA}\sim$ 6.5 km s$^{-1}$, respectively. In contrast, Fig.~\ref{fig:fig3}(b) and \ref{fig:fig3}(c) show that COF-102-2 and COF-102-3 exhibit significantly broadened phonon modes in the sub-THz region. Moreover, these structures show reduced group velocities, as shown by the softening of low-frequency acoustic phonon modes compared to COF-102-1. The observed phonon softening and band broadening in them are attributed to previously discussed increased chain rotations, as reported in prior studies \cite{ma_PCFF_chainrotation_2019, ma_poresizevsKappa2_2019}. We additionally note that, particularly in COF-102-2 and COF-102-3, bor topologies generally exhibit reduced LA group velocities in the low-frequency limit by $\sim$ 30\% compared to ctn topologies, along with smaller vibrational density of states (VDOS) peak intensity (see \SM~Sec.~SVI). We attribute our observations to bending deformations of trigonal linkers, which introduce additional vibrational scattering. Previous studies have reported that flexible C–C bonds in carbon-based structures can induce structural deformations, enhancing phonon scattering and thereby reducing phonon lifetimes and thermal conductivity, which is consistent with our observations \cite{zhang_bending_2018,zhang_bending2_2018,ying_bending3_2021}. It is also worth noting that speeds of sound in COF-102 derivatives depend on linker length, in contrast to MOFs. This behavior can be understood by applying a simple 1D chain model as discussed in Ref.~\cite{kamencek_isoMOF_phonon_2019} (see \SM~Sec.~SVII for detailed discussions).

Next, we computed the spectral MFPs following the procedure in Ref.~\cite{robbins_PCFF_crystalline_2015}. Figure~\ref{fig:fig3}(d) shows the MFP as a function of frequency for all COF-102 derivatives, computed using velocities extracted from the SED results, along with lifetimes yielded by Lorentzian line-shape fit in Eq.~\ref{eq:Lorentz}. We observe that COF-102-1 can support sub-THz vibrational modes with MFPs exceeding 1~$\mu$m, resulting from their high group velocities and long lifetimes. Noting that the specific heat of COF-102-1 is larger than the others owing to its substantially large mass density, such long MFPs, along with large group velocities, imply that these modes may significantly contribute to the observed orders of magnitude increase in thermal conductivity in COF-102-1, as also reported in $\pi$-conjugated 2D porous polymers~\cite{ma_poresizevsKappa2_2019}.

    \begin{figure}
 {\includegraphics[width=16cm, height=12cm,keepaspectratio]{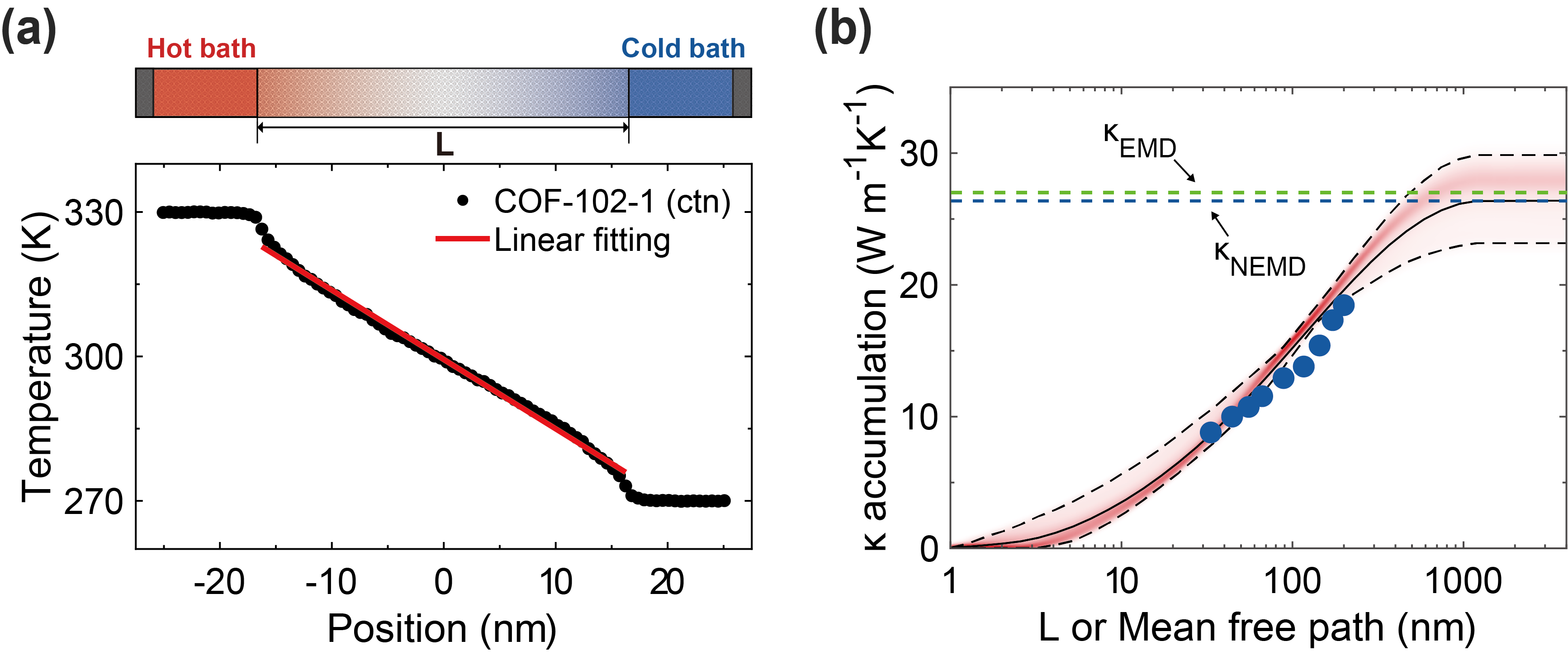}}
    \caption{\label{fig:fig4}
    (a) Illustration of NEMD configuration along with a representative temperature distribution of COF-102-1 (ctn) at 300 K. $L$ represents a characteristic length of a simulation cell (distance between thermal baths). The red line indicates the linear fit of the steady-state temperature profile, which is used for thermal conductivity calculation.
    (b) Characteristic length $L$ dependent thermal conductivity (blue dots) along with thermal conductivity accumulation versus MFPs (shaded region). Linear dashed line corresponds to bulk thermal conductivity (green: EMD results; blue: extrapolated value using NEMD results). The intensity of shaded regions (red) illustrates probability density, indicating the likelihood of the reconstructed profile of cumulative thermal conductivity. The black solid line corresponds to mean distribution enclosed by the 95\% credible interval (dashed lines). Nearly 50\% of the total thermal conductivity is contributed by phonons with MFPs above $\sim 100$ nm.}
\end{figure}

We further calculated the cumulative thermal conductivity using the non-equilibrium molecular dynamics (NEMD) method. Unlike the EMD method, NEMD predictions are strongly affected by characteristic length $L$ of simulated system, known as size effect. Figure~\ref{fig:fig4}(a) illustrates the temperature profile of COF-102-1 (ctn), along with a schematic of our simulation setup. In our NEMD simulations, briefly, two fixed regions were placed at the ends of simulation cell, and two Langevin thermostats with a 60 K temperature difference were used to generate heat flux. Finally, thermal conductivity was extracted from the best linear fit of steady-state temperature profile computed by Eq.~\ref{eq:fourier} (see Sec.~\nameref{sec:method} for details).

Figure~\ref{fig:fig4}(b) shows computed thermal conductivity versus $L$, together with cumulative thermal conductivity versus MFPs. According to NEMD simulations, the thermal conductivity of COF-102-1 (ctn) increases from 8.8 \wmk~to 18.5 \wmk~as $L$ extends from $\sim 30$ nm to $\sim 200$ nm. To estimate bulk thermal conductivity from NEMD ($\kappa_{NEMD}$), we extrapolated the $1/\kappa$ versus the $1/L$ data, yielding a value of $\sim 26.3$~\wmk, which is in a quantitative agreement with our EMD results~(see \SM~Sec.~SVIII).

We next seek to extract thermal conductivity accumulation versus MFPs using the NEMD results by solving an inverse problem described in Ref.~\cite{wei_mfpgraphite_2014}. Since the solution of an inverse problem is not unique, we reconstructed all possible solutions using the approach established in prior studies \cite{robbins_ballistic_2019, TK_aSi}. Briefly, this approach enables reconstruction of MFP distribution that satisfies smoothness criteria of priors, while reproducing simulated data by incorporating uncertainty (see \SM~Sec.~SIX for descriptions). We obtained the probability distribution of the thermal conductivity accumulation function along with the 95\% credible interval using Bayesian inference with a Metropolis--Hastings Markov chain Monte Carlo method~\cite{robbins_ballistic_2019, TK_aSi}.

The reconstructed MFP spectra are given as shaded region in Fig.~\ref{fig:fig4}(b). We observe that phonons with MFPs on the order of 100 nm contribute significantly to the heat conduction. A quantitative analysis based on the mean distribution of the MFP spectra indicates that $\sim 50$\% of the thermal conductivity is contributed by phonons with MFPs exceeding 100 nm. For acoustic phonons, modes with hundreds of nanometer scale MFP are supported up to $\sim 3$ THz, as evidenced by SED results. This finding indicates that acoustic phonons with frequencies below $\sim 3$ THz can substantially contribute to the heat conduction in COF-102-1. 

We note that the MFP distribution of COF-102-1 with significant contribution from MFPs on the order of 100 nm is similar to that in a perfect PE crystal, despite quantitatively smaller thermal conductivity of COF-102-1. More precisely, prior ab-initio calculation on a perfect PE crystal, which reported thermal conductivity of $\sim 160$~\wmk,~showed that phonon MFPs range from 1 nm to 1 $\mu$m, with $\sim 40$\% of the heat conduction carried by phonon modes with MFP $> 100$~nm, in a quantitative agreement with our results \cite{shulumba_PRL_2017}. Note that the observed contributions from MFPs $>100$ nm is also quantitatively similar to those experimentally resolved in macroscopic films; $\sim40$\% for partially ordered PE \cite{robbins_ballistic_2019} and roughly $\sim50$\% for highly ordered PE \cite{kim_originhighk_2022}.

We first discuss the role of chain rotation in the thermal conduction of organic materials. Prior MD simulations have demonstrated that chain rotation of phenyl rings tends to reduce thermal conductivity of polymers, as rotational disorder softens acoustic branches and shortens phonon MFPs \cite{ma_PCFF_chainrotation_2019}. For single-chain polymers, in particular, the impact of the chain rotation on suppressing thermal conductivity was reported to be more pronounced as the chain length increases~\cite{liu_length_dependent_2012}. In the case of bulk aromatic polymers, thermal conductivity is higher than that of their single-chain counterpart, owing to $\pi$-$\pi$ stacking between adjacent chains, which constrains random motion of phenyl rings \cite{yang_pi_pi_2023}. 
Similar to these polymeric materials, our 3D COFs with shorter phenyl-based linkers also exhibit reduced chain rotations and enhanced thermal conductivity. 
These results suggest that restricting rotation of phenyl rings provides a general design strategy for improving intrinsic thermal conductivity of organic solids.

Finally, we discuss potential strategies that could further enhance the thermal conductivity of COFs. Recent studies have highlighted the importance of phonon focusing accompanied by anisotropic bonding in polymer crystals for achieving the intrinsic upper limits of thermal conductivity in them~\cite{cheng_ab_PT_2019}. For instance, a perfect PT crystal has been shown to exhibit higher uniaxial thermal conductivity along the chain direction compared to silicon, despite its phonon lifetimes being an order of magnitude lower. This enhancement is attributed to exceptional phonon focusing effects, where high phonon group velocities are aligned along the chain axis.
Furthermore, it was demonstrated that the thermal conductivity of PT along the chain direction would drop significantly~from 197~\wmk~to 34~\wmk~if the material were to become isotropic. Supporting observations in Ref.~\cite{cheng_ab_PT_2019}, our findings suggest that the thermal conductivity of COFs has not yet reached its upper bound, as COF-102 derivatives investigated in this study are purely isotropic. We anticipate that thermally anisotropic COFs that have anisotropic bonding effect may exhibit higher thermal conductivity, which is a topic of future work.

\section{Conclusions}
In summary, we report the thermal transport properties of 3D COFs using MD simulations. Among COF-102 derivatives with varied linker lengths and topologies, we found that the derivatives with high mass density and elastic stiffness, COF-102-1, exhibit the thermal conductivity of $\sim27$ \wmk, nearly a two-fold increase relative to original COF-102. We attribute the increase in the thermal conductivity to suppressed chain rotation, leading to thermal phonons primarily scattered by anharmonicity. The computed thermal conductivity accumulation functions based on the NEMD results show that heat is predominantly conducted by low-frequency phonons with MFP exceeding 100 nm. Our work provides comprehensive insights into engineering heat conduction in COFs for thermal management applications.

\section*{Supporting Information}
Structural and physical properties of COF-102 derivatives, details of EMD thermal conductivity calculations, XRD patterns, additional PDE maps for bor structures, full phonon dispersion relations using SED analysis, effect of bending deformation on VDOS, dependence of speeds of sound, extrapolation of bulk thermal conductivity in NEMD simulations, MFP reconstruction using Bayesian inference.

\section*{Acknowledgments}
This work was supported by the National Research Foundation of Korea (NRF) grant funded by the Korea government(MSIT)(No. RS-2023-00211070) and the National Supercomputing Center with supercomputing resources including technical support (KSC-2023-CRE-0537). The authors thank Hyemin Kim for discussions.

\clearpage

\bibliographystyle{unsrtnat}
\bibliography{COFRef_revision}

\clearpage
\section*{TOC Graphic}
\begin{figure}[h]
    \centering
    \includegraphics[width=8.25cm,keepaspectratio]{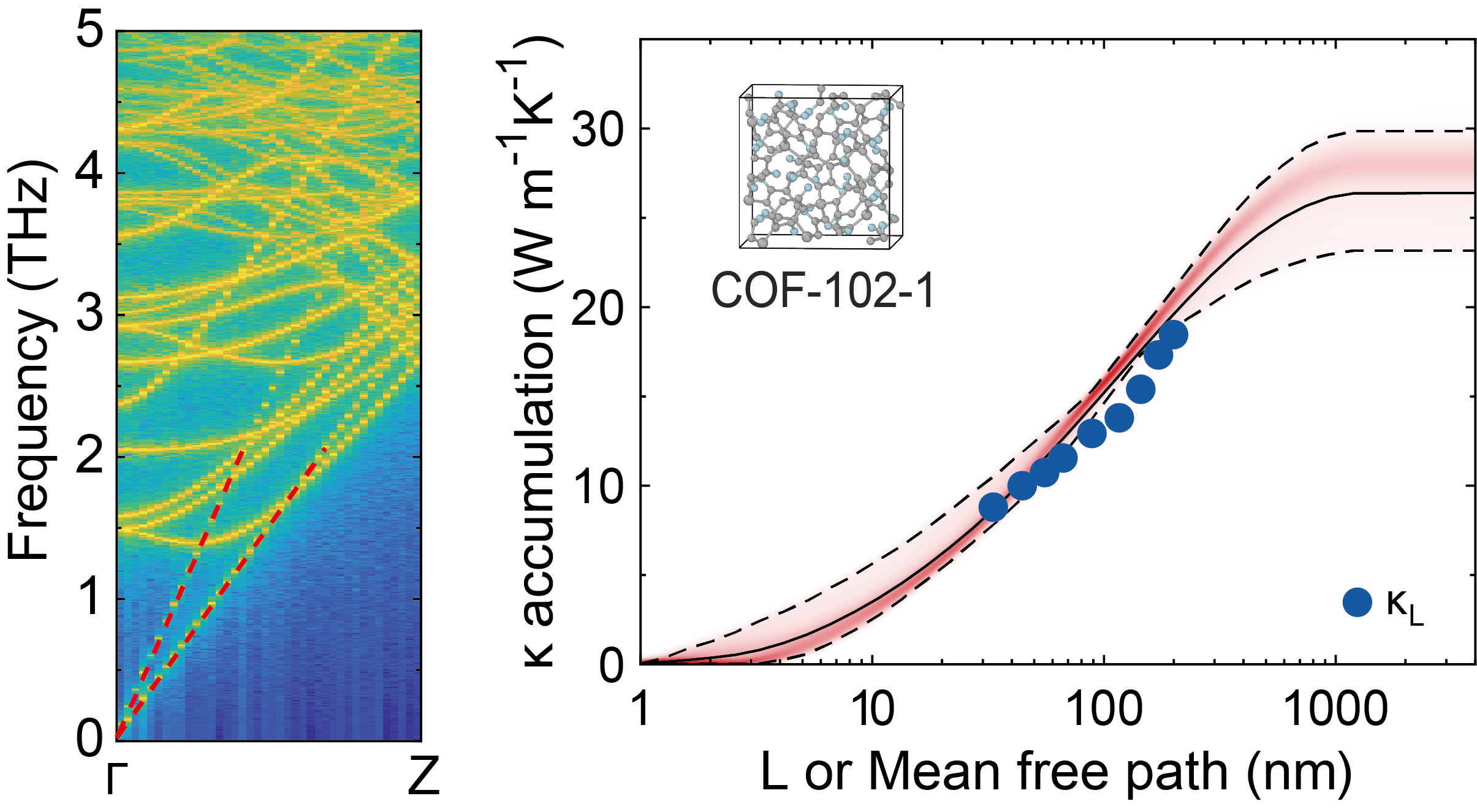}
\end{figure}

\end{document}


\title{Supporting Information: High Thermal Conductivity Dominated by Thermal Phonons with Mean Free Paths Exceeding Hundred Nanometer in Three-Dimensional Covalent Organic Framework Derivatives: A Molecular Dynamics Study}

\author{Sungjae Kim~\orcidicon{0009-0000-4062-1176}}

\affiliation{%
Department of Mechanical Engineering, Seoul National University, Seoul 08826, Republic of Korea}%

\author{Taeyong Kim~\orcidicon{0000-0003-2452-1065} }
 \email{tkkim@snu.ac.kr}
\affiliation{%
Department of Mechanical Engineering, Seoul National University, Seoul 08826, Republic of Korea}%
\affiliation{
Institute of Advanced Machines and Design, Seoul National University, Seoul 08826, Republic of Korea
}
\affiliation{%
Inter-University Semiconductor Research Center, Seoul National University, Seoul 08826, Republic of Korea}%

\date{\today}
{\global\let\newpagegood\newpage
    \global\let\newpage\relax
\maketitle}

\clearpage
\section{Structural and physical properties of COF-102 derivatives}

\setlength{\tabcolsep}{8pt} %
This section presents calculated structural and physical properties of COF-102 derivatives considered in this work. Table~\ref{Tab:phys} summarizes the chemical formula, molar mass ($M$), lattice constant ($a$), density, and pore size of each structure. The pore size was evaluated using ZEO++~\cite{willems_ZEO++_2012}.

\begin{table}[h!]
\centering
\caption{\label{Tab:phys}~Structural and physical properties of COF-102 derivatives. }

\resizebox{\textwidth}{!}{%
\begin{tabular}{ccc cccc}
\toprule
\textbf{Materials} & \textbf{Topology} & \textbf{Chemical formula} & \textbf{$M$ (g/mol)} & \textbf{$a = b = c$ (\AA)} & \textbf{Density (g/cm\textsuperscript{3})} & \textbf{Pore size (nm)} \\
\midrule
\multirow{2}{*}{COF-102-1} 
  & ctn & C\textsubscript{108}H\textsubscript{48}   & 1345.57 & 11.01 & 1.68 & 0.19 \\
  & bor & C\textsubscript{27}H\textsubscript{12}    & 336.39  & 7.13  & 1.54 & 0.41 \\
\midrule
\multirow{2}{*}{COF-102-2} 
  & ctn & C\textsubscript{396}H\textsubscript{240} & 4998.28 & 26.87 & 0.43 & 0.90 \\
  & bor & C\textsubscript{99}H\textsubscript{60}   & 1249.57 & 17.44 & 0.39 & 1.46 \\
\midrule
\multirow{2}{*}{COF-102-3} 
  & ctn & C\textsubscript{684}H\textsubscript{432} & 8650.98 & 42.59 & 0.19 & 1.79 \\
  & bor & C\textsubscript{171}H\textsubscript{108} & 2162.75 & 27.73 & 0.17 & 2.59 \\
\bottomrule
\end{tabular}}

\end{table}

\clearpage

\section{Additional description of thermal conductivity calculation using equilibrium molecular dynamics (EMD) simulations}

\begin{figure}[hbt!] 
\includegraphics[width=\textwidth,height=.4\textheight,keepaspectratio]{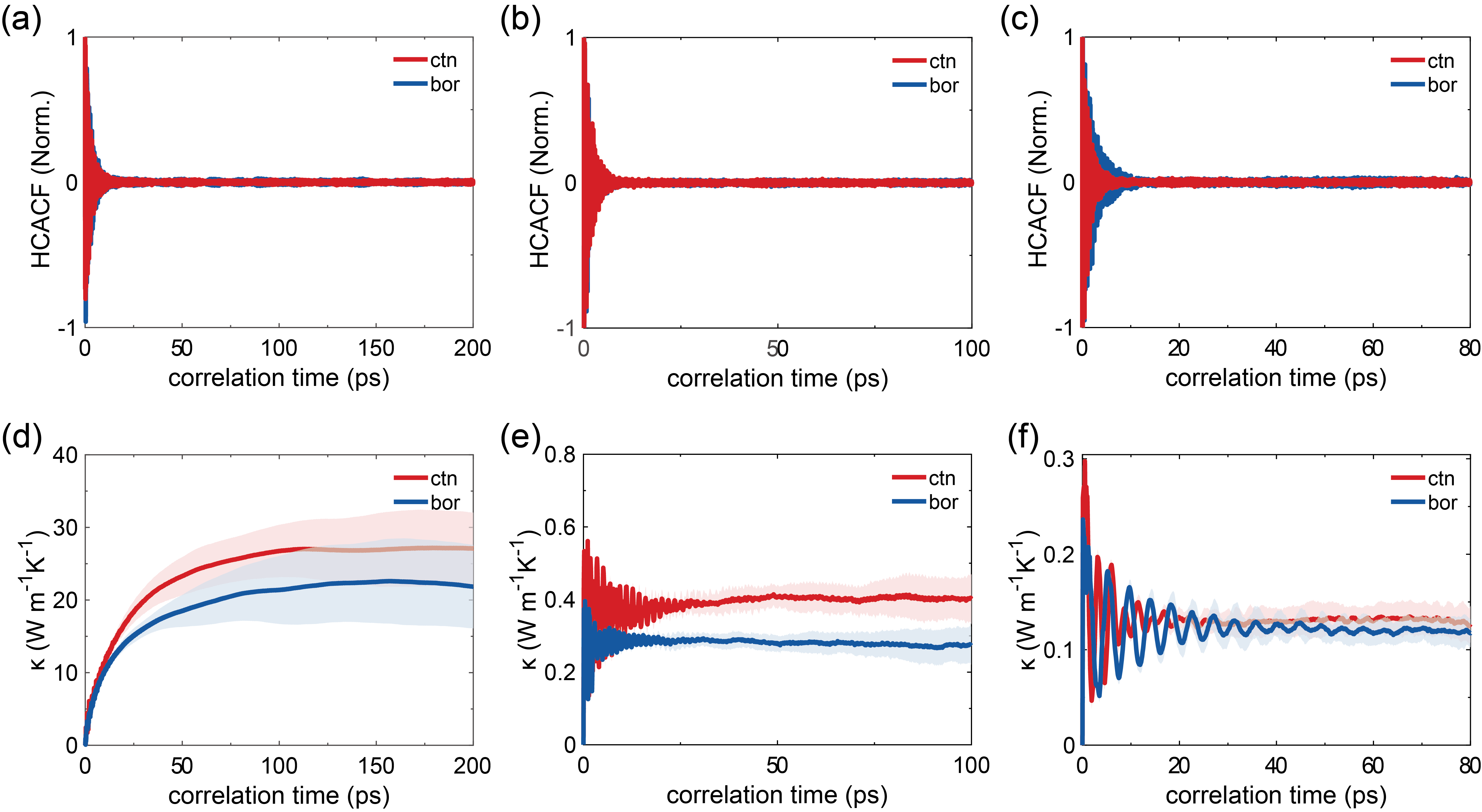}
\caption{Normalized HCACFs versus correlation time (top panels) along with the corresponding thermal conductivities (bottom panels), as computed using the Green-Kubo formalism for COF-102-1 (left), COF-102-2 (center), and COF-102-3 (right) at 300 K. For thermal conductivities, solid lines indicate average values, while shaded regions represent the standard deviations obtained from eight independent runs.}
\label{SIfig:SCfig}
\end{figure}

To compute the thermal conductivity using EMD, we used the Green-Kubo formalism as described in the main text. Figure~\ref{SIfig:SCfig} presents normalized heat current autocorrelation functions (HCACF) and the corresponding thermal conductivities of COF-102 derivatives at 300 K. In all cases, total correlation time used in our simulations exceeds 30 ps, which is sufficiently long to ensure the convergence of the thermal conductivity. To minimize statistical errors, final values were obtained by averaging results from eight independent simulations.

\begin{figure}[hbt!] 
\includegraphics[width=\textwidth,height=.25\textheight,keepaspectratio]{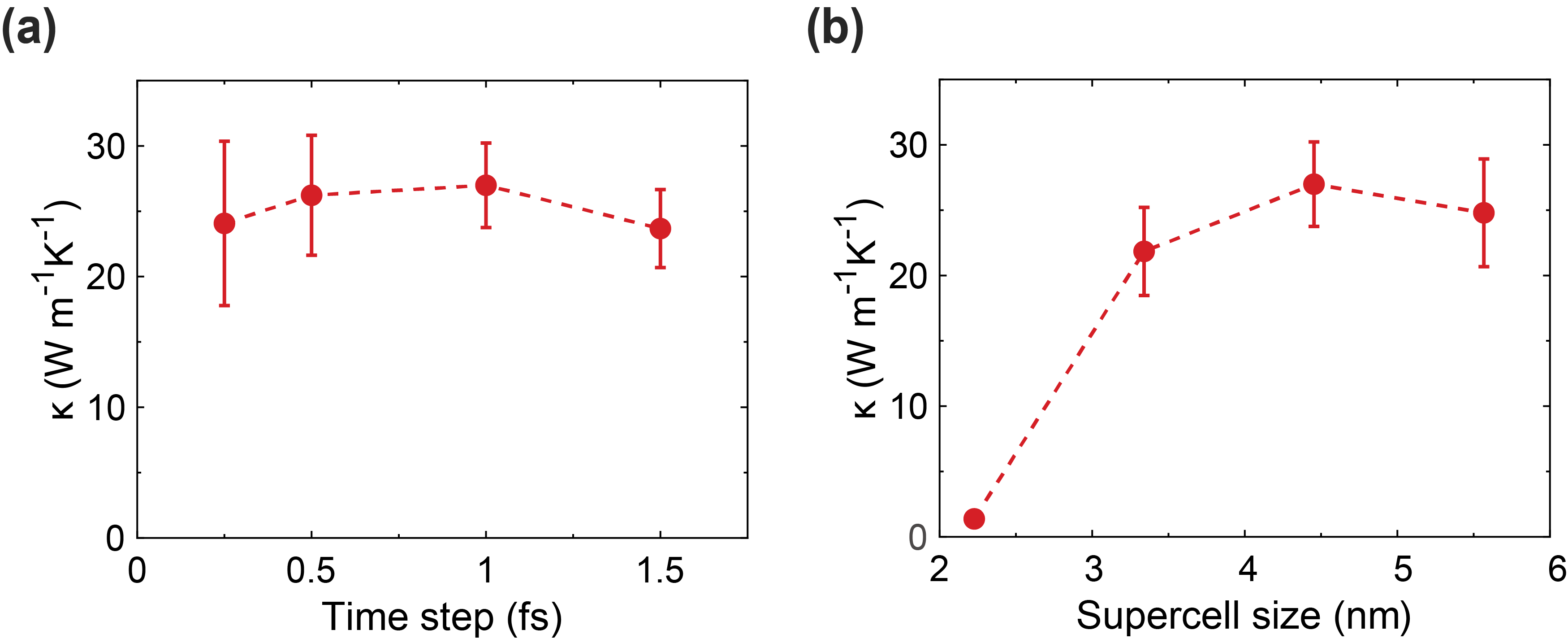}
\caption{Calculated thermal conductivity of COF-102-1 (ctn) versus (a) simulation time step and (b) supercell size. }
\label{SIfig:Bcfig}
\end{figure}

To verify that our simulation results are not influenced by simulation parameters, we calculated  thermal conductivities of COF-102-1 (ctn) using different time steps and domain sizes, as shown in Fig.~\ref{SIfig:Bcfig}. After convergence test, we chose a time step of 1 fs and a domain size of 4.45 nm as the final simulation parameters. Similar procedures were applied to the other COF structures, from which final supercell sizes were determined, as given in the method section in the main text.

\begin{figure}[hbt!] 
\includegraphics[width=\textwidth,height=.25\textheight,keepaspectratio]{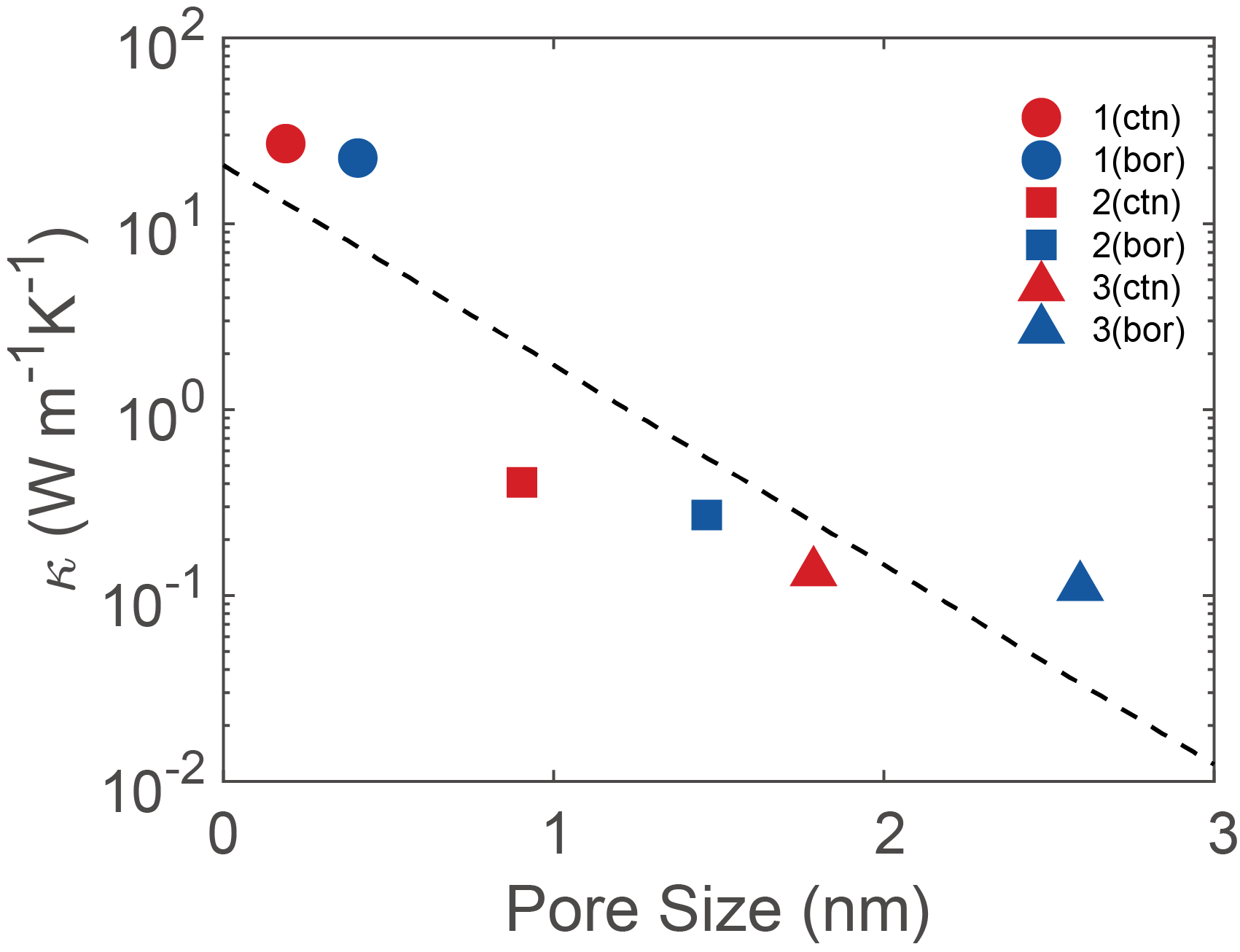}
\caption{Calculated thermal conductivities of COF-102 derivatives (red: ctn, blue: bor) as a function of their pore size.}
\label{SIfig:Porefig}
\end{figure}

Figure~\ref{SIfig:Porefig} presents calculated thermal conductivities versus pore size. The thermal conductivity decreases with increasing pore size.

\clearpage

\label{SIsec:EMD}

\clearpage
\section{X-ray diffraction (XRD) patterns}
\begin{figure}[hbt!] 
\includegraphics[width=\textwidth,height=.4\textheight,keepaspectratio]{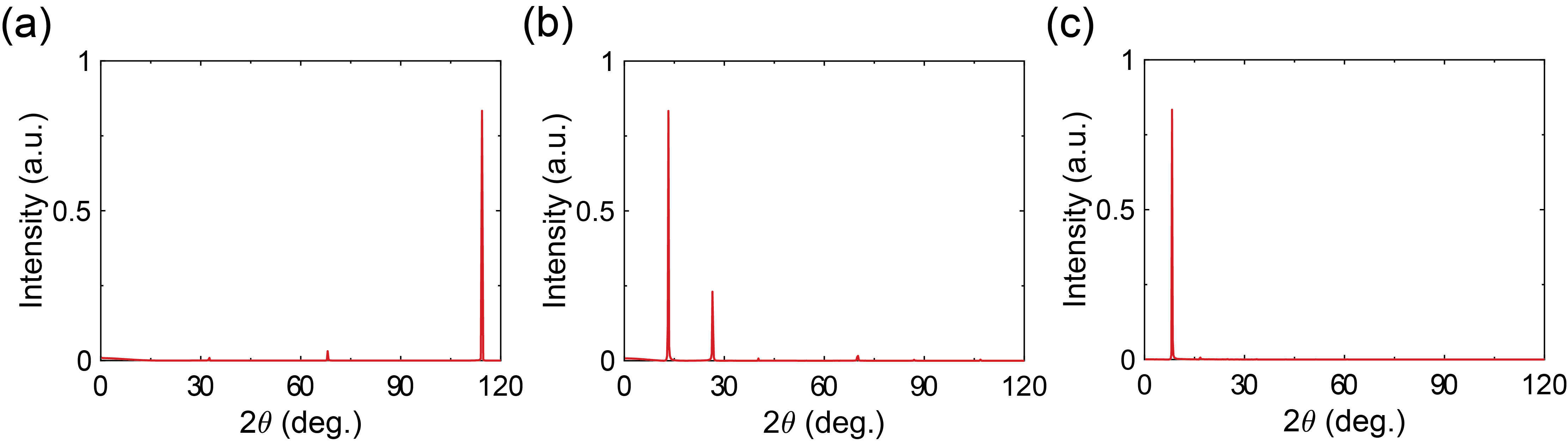}
\caption{Calculated XRD patterns of (a) COF-102-1 (ctn), (b) COF-102-2 (ctn), and (c) COF-102-3 (ctn).}
\label{SIfig:XRDfig}
\end{figure}

This section shows the XRD patterns mentioned in the main text. To evaluate the degree of crystallinity in our structures, we computed the XRD patterns using LAMMPS. The XRD intensity at each reciprocal node was calculated over the entire simulation domain size of $\sim400~\text{\AA}$, using a simulated radiation wavelength of 1.54~\text{\AA}. Figure~\ref{SIfig:XRDfig} shows the resulting XRD patterns, all of which exhibit sharp peaks, indicating a high degree of crystallinity.

\clearpage

\section{Additional normalized probability density estimate (PDE) maps for bor structures}

\begin{figure}[hbt!] 
\includegraphics[width=\textwidth,height=.4\textheight,keepaspectratio]{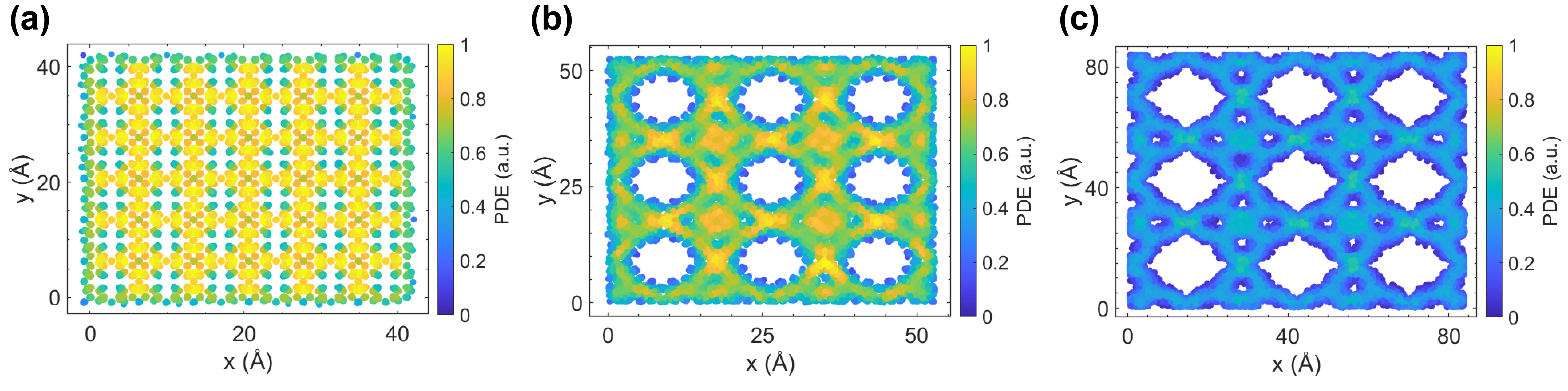}
\caption{Normalized PDE maps for simulated structures of (a) COF-102-1 (bor), (b) COF-102-2 (bor), and (c) COF-102-3 (bor)}
\label{SIfig:PDEfig}
\end{figure}

This section presents additional normalized PDE maps for bor structures as given in Fig.~\ref{SIfig:PDEfig}. Upon comparison of PDE maps, both bor and ctn topologies exhibit similar trends across the derivatives. More precisely, we confirm that the PDE also progressively decreases from COF-102-1 (bor) to COF-102-3 (bor), consistent with our result for ctn topologies presented in the manuscript.

\clearpage

\section{Additional spectral energy density (SED) representing phonon dispersion relations}
\label{SIsec:phonon_disp}
\begin{figure}[hbt!] 
\includegraphics[width=\textwidth,height=.4\textheight,keepaspectratio]{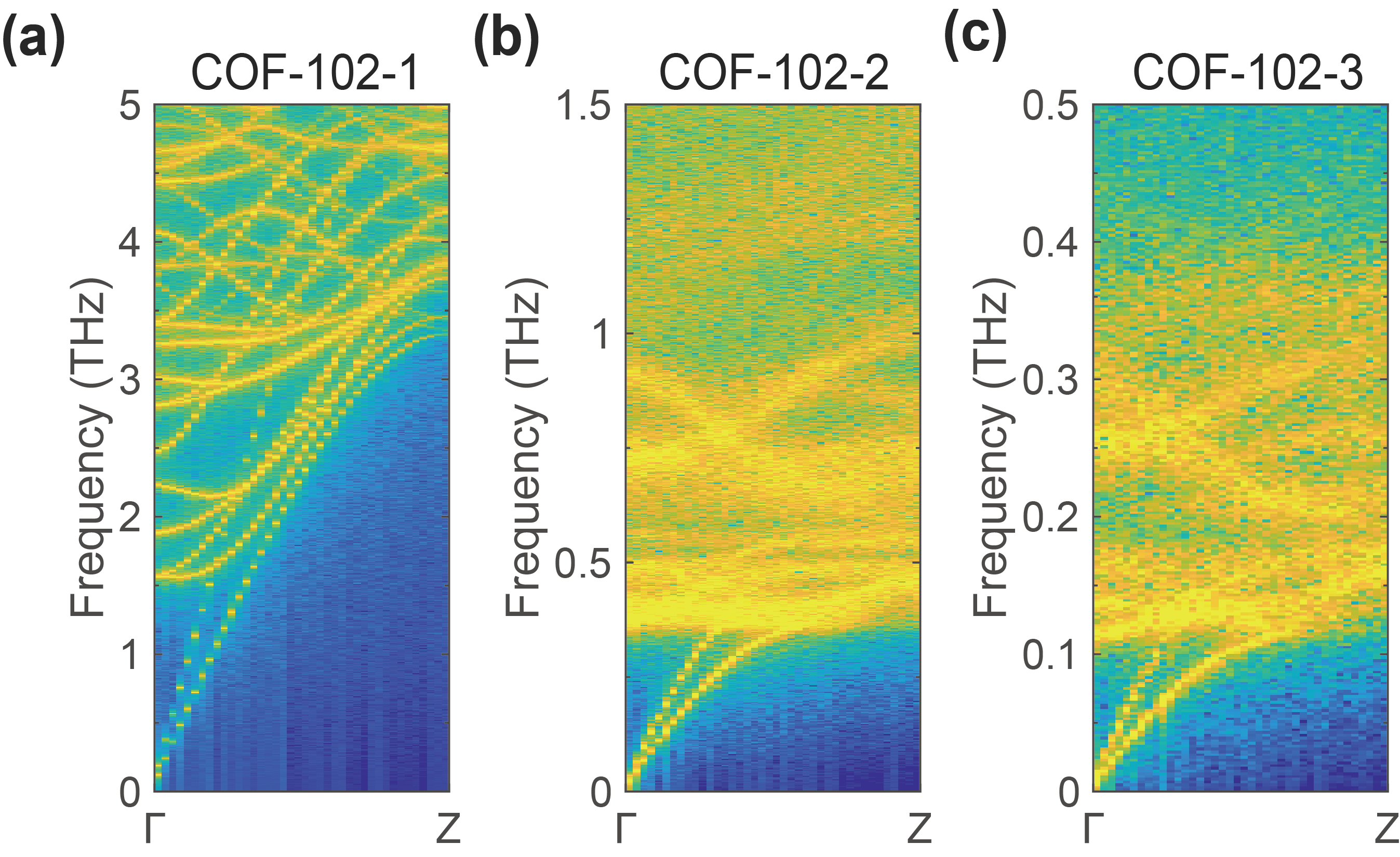}
\caption{SED over low frequency ranges, representing phonon dispersion relations at 300 K for (a) COF-102-1 (bor), (b) COF-102-2 (bor), and (c) COF-102-3 (bor)}
\label{SIfig:SEDborfig}
\end{figure}

\begin{figure}[hbt!] 
\includegraphics[width=\textwidth,height=.8\textheight,keepaspectratio]{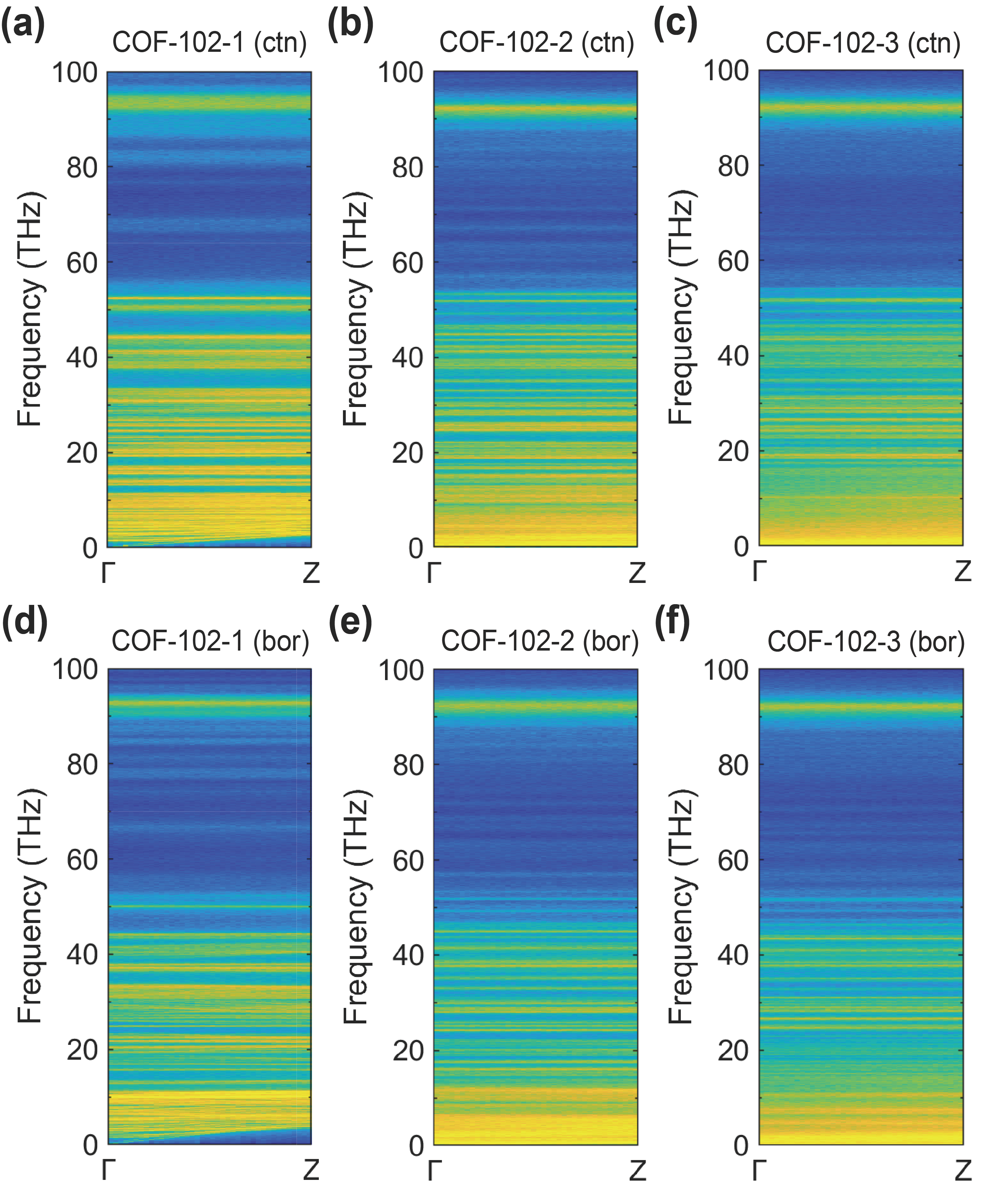}
\caption{SED over the entire frequency range, representing phonon dispersion relations at 300 K for COF-102 derivatives. Results are shown for (a) COF-102-1 (ctn), (b) COF-102-2 (ctn), (c) COF-102-3 (ctn), (d) COF-102-1 (bor), (e) COF-102-2 (bor), and (f) COF-102-3 (bor).}
\label{SIfig:fullSED}
\end{figure}

Figure~\ref{SIfig:SEDborfig} shows the phonon dispersion relations of bor structures over selected frequencies, as computed using SED analysis. The full phonon dispersion relations for all derivatives considered in this work are shown in Fig.~\ref{SIfig:fullSED}.

\clearpage

\section{Effect of bending deformation on vibrational density of states (VDOS)}

\begin{figure}[hbt!] 
\includegraphics[width=\textwidth,height=.4\textheight,keepaspectratio]{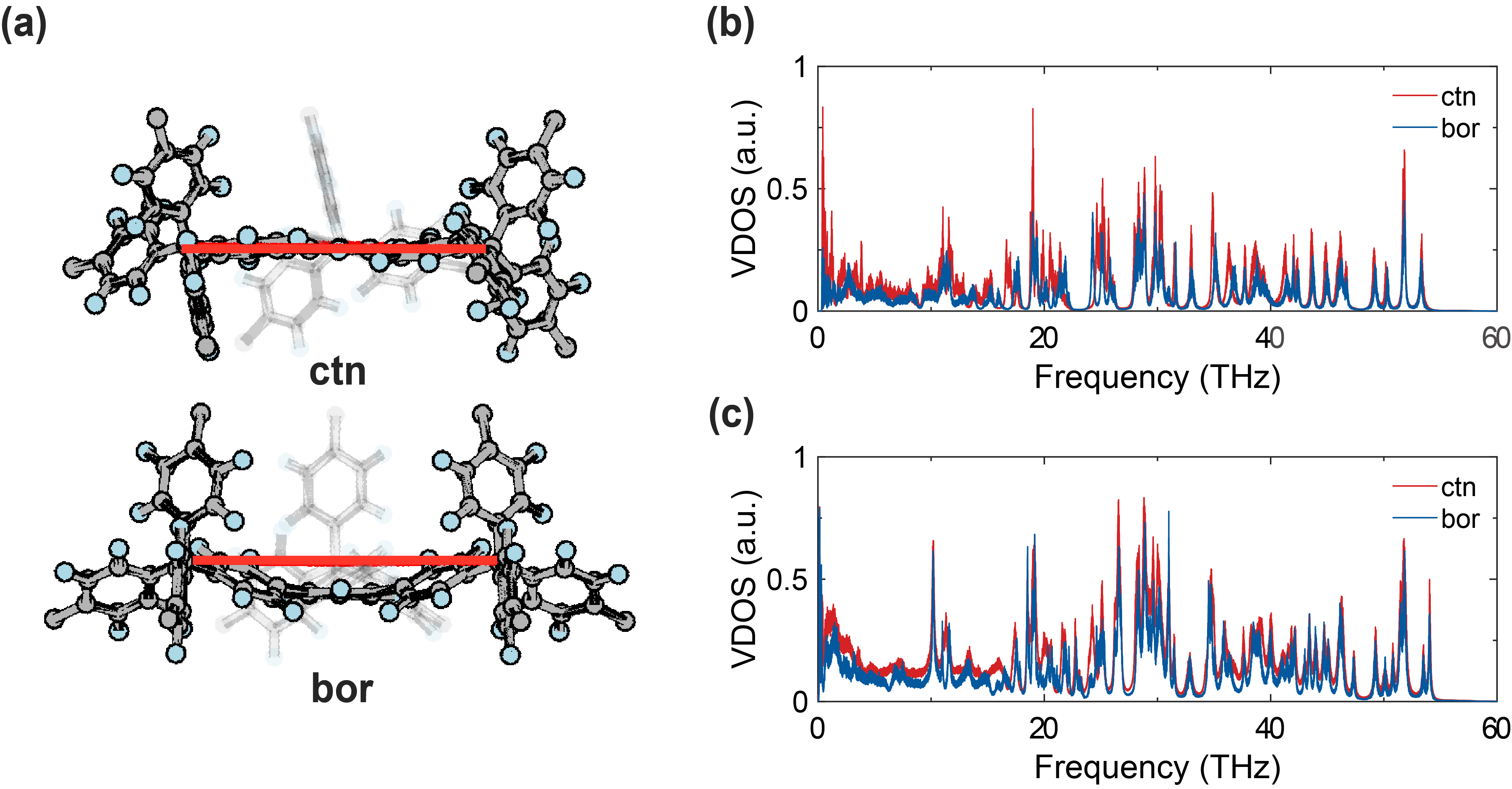}
\caption{(a) Representative atomic configurations of trigonal linkers in COF-102 derivatives. The red line represents the edge view of a plane connecting the three adjacent nodes, where a marked bending deformation is observed in bor topology. Calculated VDOS of (b) COF-102-2, and (c) COF-102-3 for both ctn and bor topologies. In general, VDOS peak intensities are reduced in bor topologies compared to ctn topologies.}
\label{SIfig:VDOS}
\end{figure}

This section presents the calculation of VDOS, from which we examined the impact of bending deformation in linkers on the vibrational properties of COFs. Briefly, VDOS was obtained by calculating the Fourier transform of atomic velocity autocorrelation function (VACF) \cite{dickey_VDOS_1969},
\begin{equation}
\mathrm{VDOS}(\omega) = \lim_{\tau \to \infty} \int_{-\tau}^{\tau} 
\gamma(t)
\, e^{-i \omega t} \, dt
\label{seq:VDOS}
\end{equation}
where $\omega$ is vibrational frequency, and 
$\gamma(t) = \left\langle \sum m_i\, \vec{v}_i(t) \cdot \vec{v}_i(0) \right\rangle / 
\left\langle \sum m_i\, \vec{v}_i(0) \cdot \vec{v}_i(0) \right\rangle$
is the VACF. In Eq.~\ref{seq:VDOS}, $m_i$ is mass of $i$-th atom, $\vec{v}_i(t)$ is velocity vector of the $i$-th atom at a time $t$. The results are given in Fig.~\ref{SIfig:VDOS}. Compared to ctn topologies, bor topologies generally exhibit reduced VDOS peak intensities, which we attribute to bending deformation in trigonal linkers.

\clearpage

\section{Dependence of speeds of sound in COF-102 derivatives}

\begin{figure}[hbt!] 
\includegraphics[width=\textwidth,height=.45\textheight,keepaspectratio]{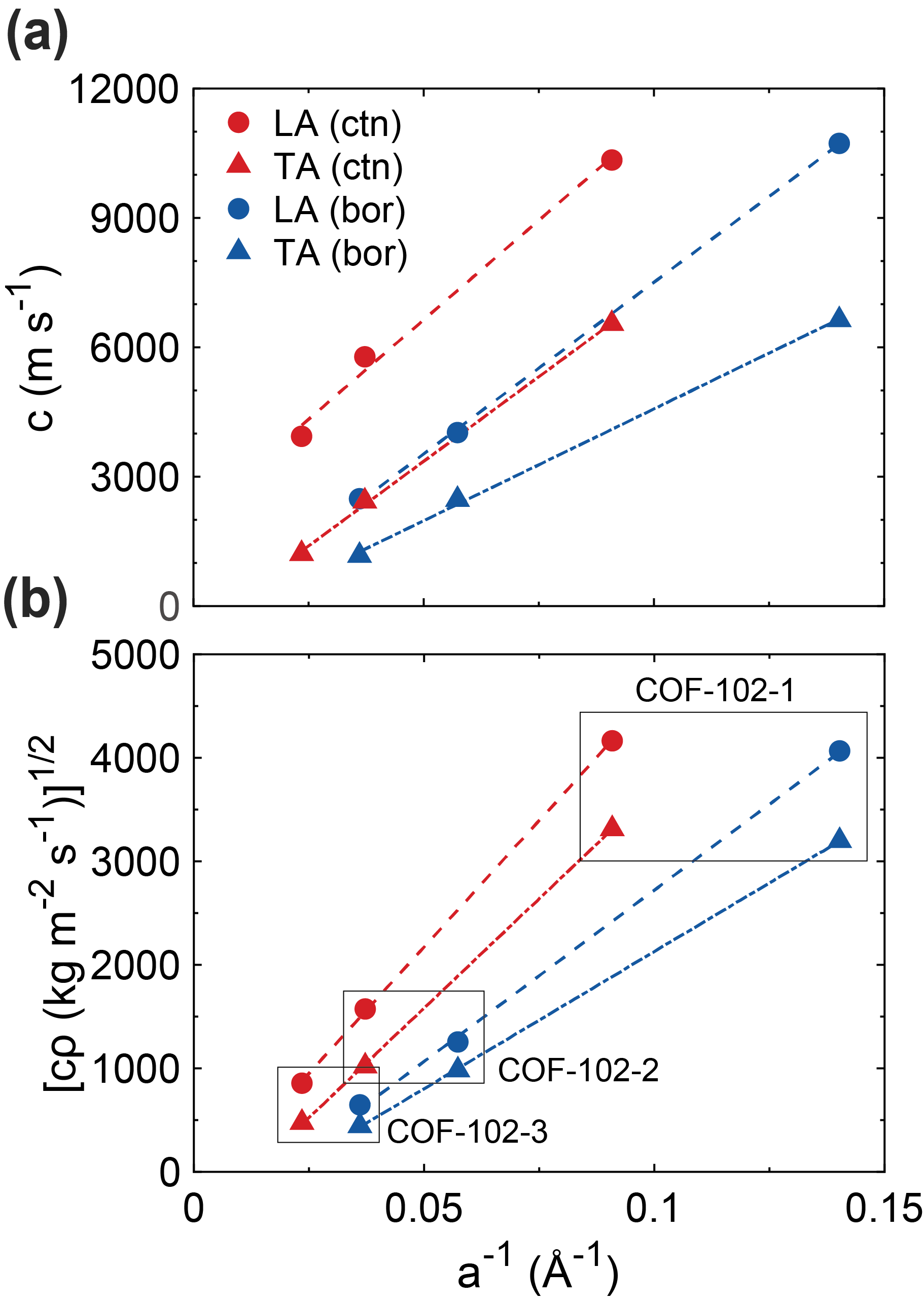}
\caption{(a) Speed of sound of acoustic branches versus inverse lattice constant ($a^{-1}$). (b) square roots of product between speed of sound and mass density versus $a^{-1}$. Note that circles (upward pointing triangles) correspond to LA (TA) mode, while red colors (blue colors) are for ctn (bor) topologies. In general, linear trend can be seen.}
\label{SIfig:SoS}
\end{figure}

In this section, we discuss the dependence of the speed of sound on linker length in COF-102 derivatives by analyzing trends observed in Fig.~\ref{SIfig:SoS} using a 1D chain model \cite{kittel_solidphysics_2018}. 

Figure~\ref{SIfig:SoS}(a) presents the speed of sound for acoustic phonon modes as a function of inverse lattice constant $a^{-1}$. As the linker length increases, additional phenyl rings are incorporated between adjacent carbon nodes, resulting in a larger lattice constant $a$ and lower mass density. This, in turn, leads to a decrease in the speed of sound in our COF-102 derivatives, in contrast to MOFs, where such changes have negligible effects on the speed of sound \cite{kamencek_isoMOF_phonon_2019}. 

To explain the observed behavior in our COFs, we employ a 1D chain model following prior literature \cite{kamencek_isoMOF_phonon_2019}, in which the speed of sound $c$ is given by
\begin{equation} \label{eq:S2}
c = a \sqrt{\gamma/m},
\end{equation}
where $a$, $\gamma$, and $m$ are lattice constant, stiffness, and mass, respectively. 

Subsequently, we multiply both sides of the Eq.~\ref{eq:S2} by the mass density $\rho= m/a^3$, which leads to the following expression:
\begin{equation} \label{eq:S3}
c \rho = \sqrt{\gamma m}\frac{1}{a^2}.
\end{equation}

Figure~\ref{SIfig:SoS}(b) shows $(c\rho)^{1/2}$ versus $a^{-1}$, obtained by taking the square root of Eq.~\ref{eq:S3}. A clear linear trend is observed, suggesting that $\sqrt{\gamma m}$ remains constant across all structures. This indicates that an increase in mass due to longer linkers is compensated by a decrease in stiffness.

Next, examining Eq.~\ref{eq:S2}, the increase in $m$ resulting from longer organic linkers is accompanied by a decrease in $\gamma$, leading to a reduction in $\sqrt{\gamma/m}$; however, in MOFs, this effect is offset by a simultaneous increase in lattice constant $a$ that rivals $\sqrt{\gamma/m}$, resulting in a nearly constant speed of sound, as observed in Ref.~\cite{kamencek_isoMOF_phonon_2019}. In contrast, the presence of lightweight organic linkers and nodes in COF-102 derivatives leads to more pronounced changes in both mass and stiffness with increasing linker length. Consequently, the significant reduction in $\sqrt{\gamma/m}$ outweighs the increase in $a$, leading to a net decrease in speed of sound. 
Thus, thermal conductivity in COFs can be readily tuned by modulating linker length, while in MOFs, the presence of heavy metal nodes causes strong mass mismatch with linkers, which limits the impact of linker length on thermal transport and thereby reduces the extent of tunability.

\clearpage
\section{Extrapolation of bulk thermal conductivity in NEMD simulations}

\begin{figure}[hbt!] 
\includegraphics[width=\textwidth,height=.25\textheight,keepaspectratio]{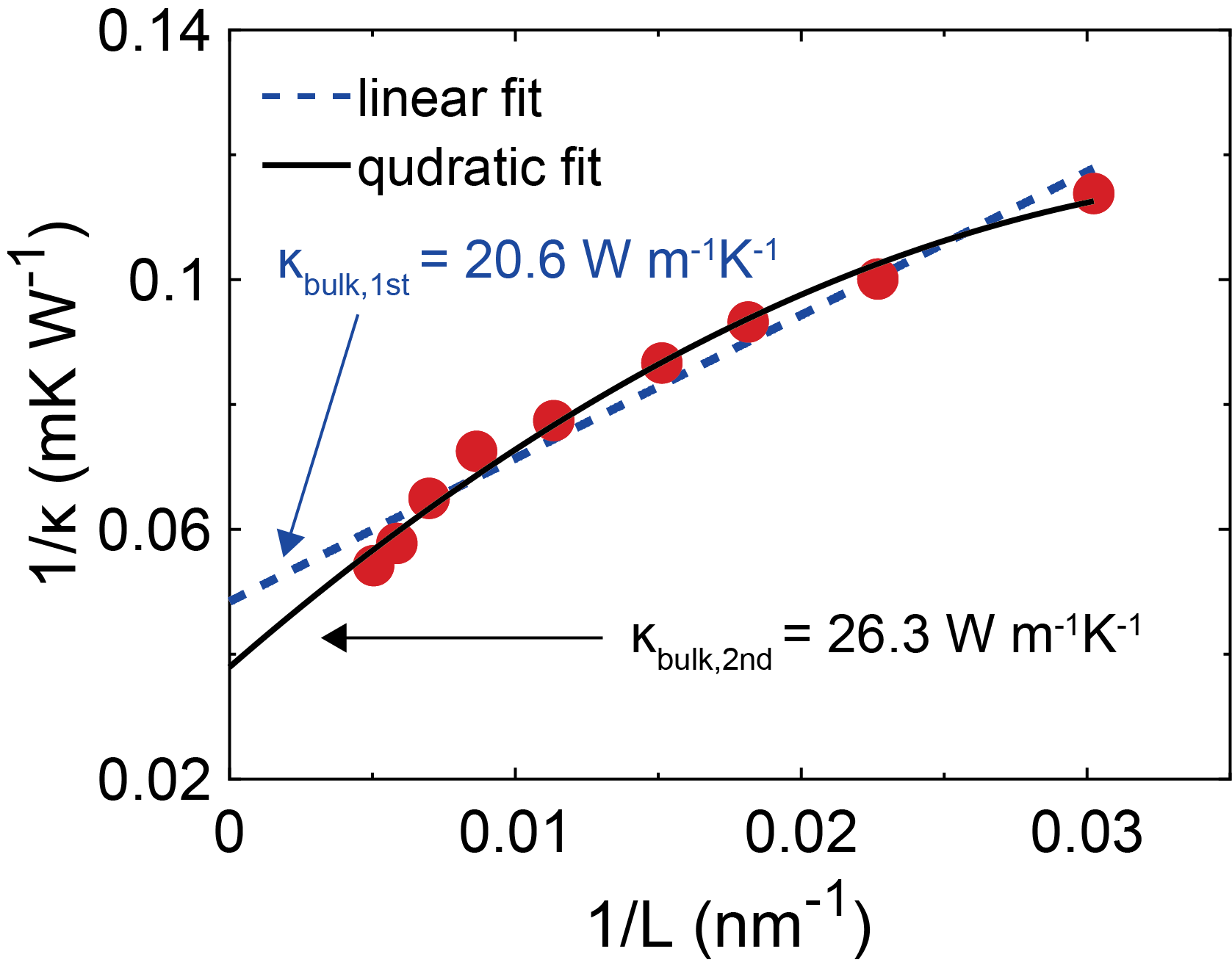}
\caption{Calculated inverse of thermal conductivity versus inverse of characteristic lengths $L$ in NEMD simulations for COF-102-1 (ctn), represented as red markers.
The solid black curve represents the quadratic fit using Eq.~\ref{eq:S5} to our data that are compatible with simulation results, while the dashed blue line shows the linear fit using Eq.~\ref{eq:S4} which fails to capture the observed trend. 
\label{SIfig:Extrapo}}
\end{figure}

We determined the thermal conductivity at $L \rightarrow \infty$, in which $L$ is defined as characteristic length in NEMD simulations. Figure~\ref{SIfig:Extrapo} shows the inverse of thermal conductivity as a function of the inverse of $L$, exhibiting a nonlinear trend as observed in prior literature~\cite{talaat_extrapolation_2020,dong_SizeNEMD_2018}.
We first estimated the bulk thermal conductivity by linearly extrapolating to the limit 1/$L$ $\rightarrow$ 0 using the following relation,
\begin{equation}
\frac{1}{\kappa(L)} = \frac{1}{\kappa_{\text{bulk}}} \left(1 + \frac{\lambda}{L} \right),
\label{eq:S4}
\end{equation}
where $\lambda$ is effective phonon MFP. However, this linear extrapolation yields a bulk thermal conductivity of 20.6~\wmk, which fails to capture the nonlinear trend observed in simulation data. 
Instead, we used second-order extrapolation of 
\begin{equation}
\frac{1}{\kappa(L_)} = \frac{1}{\kappa_{bulk}} \left(1 + \frac{\lambda}{L} + \frac{\beta}{L^2} \right),
\label{eq:S5}
\end{equation}
where $\beta$ represents an arbitrary constant. This approach yields a bulk thermal conductivity of 26.3~\wmk, which quantitatively agrees with our EMD results ($\sim$ 27~\wmk).

\clearpage

\section{MFP reconstruction using Bayesian inference}

To reconstruct the phonon MFP accumulation function $F(\Lambda)$ from NEMD simulations, we solve the following equation \cite{wei_mfpgraphite_2014}:
\begin{equation}
\kappa(L) = \int_0^\infty F(\Lambda) \cdot \frac{12L}{(4\Lambda + 3L)^2} \, d\Lambda.
\label{eq:kap}
\end{equation}
Here, $\kappa(L)$ is the thermal conductivity obtained using NEMD simulations, which depends on characteristic length $L$, and $\Lambda$ is phonon MFP. The kernel function ${12L}/{(4\Lambda + 3L)^2}$ represents the suppression of effective MFP due to boundary scattering at finite $L$. As discussed in the main text, the solution of the ill-posed problem of Eq.~\ref{eq:kap} is known to be not unique.

In this work, we used Bayesian inference scheme that incorporates uncertainty in the simulation data to extract distributions of possible solutions, as is widely used in prior experimental studies~\cite{ravichandran_spectrally_2018, robbins_ballistic_2019}.
We followed the approach employed in Refs.~\cite{ravichandran_spectrally_2018, robbins_ballistic_2019, Chen_MFPspectroscopy_2020} in which details of the implementations can be found. Briefly, we estimated the credible interval for $F(\Lambda)$ using a Metropolis--Hastings Markov chain Monte Carlo (MHMCMC) method, where the posterior probability of $F(\Lambda)$ is evaluated as
\begin{equation}
P_{\text{post}}(F \mid \kappa_{\text{sim}}) \propto P(\kappa_{\text{sim}} \mid F) \cdot P_{\text{prior}}(F).
\label{eq:S7}
\end{equation}
where $\kappa_{sim}$ is the simulation data set obtained using NEMD. Noting that $P(\kappa_{\text{sim}} \mid F)$ in Eq.~\ref{eq:S7} is defined by a normal distribution $\mathcal{N}(\kappa_{\text{sim}}-\kappa_F, \, \sigma^2 I)$, it corresponds to the likelihood that quantifies how well $\kappa_F$, the thermal conductivity predicted by the trial $F(\Lambda)$, reproduces $\kappa_{sim}$, as evaluated using the standard deviation ($\sigma$) from the NEMD results. Given our prior $F$, which we impose smoothness criteria, the prior probability $P_{\text{prior}}(F)$ is modeled as $\mathcal{N}(\Delta^2 F, \, \gamma^2 I)$, where $\gamma$ is the parameter that we perturbed to adjust smoothness.

\clearpage

\bibliographystyle{unsrtnat}
\bibliography{COFRef_revision}